\documentclass[a4paper,12pt]{article}
\usepackage{graphicx}

\newtheorem{definition}{{\bf Definition}}[section]
\newtheorem{theorem}[definition]{{\bf Theorem}}

\newtheorem{corollary}[definition]{{\bf Corollary}}

\def\<{\langle}
\def\>{\rangle}
\def\im{{\rm i}}

\def\Hc{{\cal H}}

\def\proof{\noindent{Proof. }}
\def\endproof{\hfill $\square$ \vspace{10pt}}
\usepackage{amscd,amsmath,amssymb,graphicx,cite}

\begin{document}

\title{Unitary equivalent classes of one-dimensional quantum walks II}
\author{Hiromichi Ohno  \medskip \\
Department of Mathematics, Faculty of Engineering, Shinshu University,\\
4-17-1 Wakasato, Nagano 380-8553, Japan}
\date{}

\maketitle

\begin{abstract}
This study investigated the unitary equivalent classes of one-dimensional quantum walks.
We determined the unitary equivalent classes of one-dimensional quantum walks,
two-phase quantum walks with one defect, complete two-phase quantum walks,
one-dimensional quantum walks with one defect and translation-invariant 
one-dimensional quantum walks.
The unitary equivalent classes of  one-dimensional quantum walks 
with initial states were also considered.
\end{abstract}

\section{Introduction}

This study investigated the unitary equivalent classes of one-dimensional quantum walks.
A quantum walk is defined by a pair $(U, \{\Hc_v\}_{v \in V})$, in which
$V$ is a countable set, $\{\Hc_v\}_{v \in V}$ is a family of separable Hilbert spaces,
and $U$ is a unitary operator on $\Hc = \bigoplus_{v \in V} \Hc_v$ \cite{SS2}.
In this paper, we discuss one-dimensional (two-state) quantum walks, in which
$V = {\mathbb Z}$ and $\Hc_v = {\mathbb C}^2$.
These have been the subject of many previous studies
\cite{ADZ, ABNVW, CGMV, EEKST1, EEKST2, EK, EK2, EKST, EKO, GKD, G, KRBD, K1, K2, K3, KLS, O, SK, V, WLK}.

It is important to clarify 
what it means to say that
two quantum walks are the same.
We define unitary equivalence in the same way as \cite{SS2, O}.
If two quantum walks are unitary equivalent, 
then their digraphs, dimensions of their Hilbert spaces, and
the probability distributions of the quantum walks are the same.

In the previous paper \cite{O},
we discussed some general properties of unitary equivalent quantum walks.
In particular, we proved that every one-dimensional quantum walk is 
the unitary equivalent of one of the Ambainis type.
We also presented the
necessary and sufficient condition for defining a one-dimensional quantum walk as a Szegedy walk.

Unitary equivalent classes of simple quantum walks 
have been shown to be 
parameterized by a single parameter \cite{GKD}.
In contrast, there are several types of one-dimensional quantum walks,
including two-phase quantum walks with one defect \cite{EEKST1,EEKST2}, 
complete two-phase quantum walks \cite{EKO}, and
one-dimensional quantum walks with one defect 
\cite{CGMV, EK, EK2, EKST, K3, KLS,WLK}.
In this study, we clarified the unitary equivalent classes of 
general one-dimensional quantum walks and of the above types of one-dimensional quantum walk,
but excluding certain special cases.
In Sect. \ref{sec2},
we present our results: 
two-phase quantum walks with one defect are parameterized by six parameters,
complete two-phase quantum walks by four parameters, and 
one-dimensional quantum walks with one defect by four parameters.

When studying the probability distribution of a quantum walk,
an initial state must be set and a quantum walk with an initial state must also be considered.
In Sect. \ref{sec3},
we present unitary equivalent classes of 
all the above types of one-dimensional quantum walk with an initial state.


\section{Unitary equivalent classes of one-dimensional quantum walks}\label{sec2}

We first consider the definition of a one-dimensional quantum walk and 
the unitary equivalence of such walks (see \cite{O}).

\begin{definition}
Let $\Hc_n = {\mathbb C}^2$ for $n \in {\mathbb Z}$.
A unitary operator $U$ on $\Hc = \bigoplus_{n\in{\mathbb Z}} \Hc_n$ is
called a one-dimensional quantum walk if 
\[
{\rm rank} P_n U P_m = 
\left\{
\begin{array}{ll}
1 \quad & (m = n\pm 1) \\
0 & (m \neq n \pm 1)
\end{array}
\right. 
\]
for all $m, n \in {\mathbb Z}$, where $P_n$ is the projection onto $\Hc_n$.
\end{definition}

A (pure) quantum state is represented by a unit vector in a Hilbert space.
For $\lambda \in {\mathbb R}$,
quantum states $\xi$ and $e^{\im \lambda} \xi$ in $\Hc$ are identified. 
Hence, the one-dimensional quantum walks $U$ and $e^{\im \lambda} U$
are also identified.

\begin{definition}
One-dimensional quantum walks $U_1$ and $U_2$ are unitary equivalent if 
there exists a unitary $W = \bigoplus_{n\in{\mathbb Z}} W_n$ on 
$\Hc = \bigoplus_{n\in{\mathbb Z}} \Hc_n$ such that
\[
WU_1 W^* = U_2.
\]
\end{definition}

Theorem 1 in \cite{O} (see also the first paragraph of Section 5 in \cite{O})
yields the next theorem.

\begin{theorem}
A one-dimensional quantum walk $U$ is described as follows:
\begin{equation}\label{eq:unitary}
U =\sum_{n \in {\mathbb Z}} 
|\xi_{n-1,n}\>\< \zeta_{n-1,n}| + |\xi_{n+1,n}\>\<\zeta_{n+1,n}|,
\end{equation}
where $\{\xi_{n,n+1},\xi_{n+1,n}\}_{n\in {\mathbb Z}}$ and
 $\{\zeta_{n,n+1} ,\zeta_{n+1,n}\}_{n\in{\mathbb Z}}$ are orthonormal bases of 
$\Hc = \bigoplus_{n\in{\mathbb Z}} \Hc_n$ with $\xi_{n,n+1}, \zeta_{n+1,n} \in \Hc_n$ and
$\xi_{n+1,n},\zeta_{n,n+1} \in \Hc_{n+1}$.
\end{theorem}

The unitary equivalence of a one-dimensional quantum walk
can then be analyzed as follows.

\medskip 
\noindent
{\bf Step 1.}
Assume that $U$ is described as in \eqref{eq:unitary},
and define a unitary operator $W_1$ on $\Hc$ as
\[
W_1 = \bigoplus_{n \in {\mathbb Z}} |e_1^n\>\<\xi_{n, n+1}| + |e_2^n \>\< \xi_{n,n-1}|,
\]
where $\{e_1^n, e_2^n\}$ is the standard basis of $\Hc_n = {\mathbb C}^2$.
Then,
\begin{eqnarray}
W_1 U W_1^* &=& 
\sum_{n \in {\mathbb Z}} 
|W_1 \xi_{n-1,n}\>\< W_1 \zeta_{n-1,n}| + |W_1 \xi_{n+1,n}\>\<W_1\zeta_{n+1,n}|\nonumber \\
&=&
\sum_{n \in {\mathbb Z}} |e_1^{n-1} \>\< e^{\im a_n} r_n e_1^n + e^{\im b_n} s_n e_2^n|
+ |e_2^{n+1}\>\<e^{\im c_n} s_n e_1^n + e^{\im d_n} r_n e_2^n| \label{eq:1.1.1}
\end{eqnarray}
for some $0 \le r_n \le 1$ and $a_n, b_n, c_n, d_n \in {\mathbb R}$
with $s_n = \sqrt{1-r_n^2}$ and
\begin{equation}\label{eq:1.1.3}
a_n -b_n = c_n-d_n + \pi  \quad ({\rm mod} \ 2\pi).
\end{equation}
If there is no confusion,
$({\rm mod}\ 2\pi)$ can be omitted hereafter.

\medskip 
\noindent
{\bf Step 2.}
Define $g_n \in {\mathbb R}$ by $g_0 =0 $ and
\[
g_{n-1} - g_n = a_n,
\]
inductively. Similarly, define $h_n \in {\mathbb R}$ by $h_0 = g_{-1}-b_0$ and
\[
h_{n+1} - h_n = d_n,
\]
inductively. Then, by \eqref{eq:1.1.3},
\begin{equation}\label{eq:1.1.4}
c_n - h_{n+1}+g_n = c_n -h_n-d_n +g_{n-1} -a_n 
=-(b_n - g_{n-1} +h_n) +\pi \quad ({\rm mod} \ 2\pi).
\end{equation}
Let $W_2$ be a unitary operator defined by
\[
W_2 = \bigoplus_{n \in{\mathbb Z}} e^{\im g_n} |e_1^n\>\<e_1^n| + 
e^{\im h_n} |e_2^n\>\<e_2^n|.
\]
By definitions and \eqref{eq:1.1.4},
\begin{eqnarray*}
&& W_2 W_1 U W_1^* W_2^* \\
&=& \sum_{n \in {\mathbb Z}} |e_1^{n-1} \>\< e^{\im(a_n- g_{n-1}+g_n)} r_n e_1^n 
+ e^{\im (b_n-g_{n-1} + h_{n})} s_n e_2^n|\\
&& + |e_2^{n+1}\>\<e^{\im (c_n-h_{n+1} + g_n)} s_n e_1^n 
+ e^{\im (d_n -h_{n+1} + h_n)} r_n e_2^n| \\
&=&
\sum_{n \in {\mathbb Z}} |e_1^{n-1} \>\< r_n e_1^n 
+ e^{\im k_n} s_n e_2^n| + |e_2^{n+1}\>\<- e^{-\im k_n} s_n e_1^n 
+ r_n e_2^n|,
\end{eqnarray*}
where $k_n = b_n - g_{n-1} + h_n$.
Here, 
\[
k_0 =  b_0 -g_{-1} +h_0 = 0.
\]

\medskip 
\noindent
{\bf Step 3.}
Let $\ell = k_1/2$, $p_n = n k_1/2$ and $q_n = -n k_1/2$, and let
\[
W_3 = \bigoplus_{n \in {\mathbb Z}}
 e^{\im p_n} |e_1^n\>\<e_1^n| + 
e^{\im q_n} |e_2^n\>\<e_2^n|.
\]
Then,
\begin{eqnarray*}
&& e^{\im \ell} W_3 W_2 W_1 U W_1^* W_2^* W_3^* \\
& =& 
\sum_{n \in {\mathbb Z}} |e_1^{n-1} \>\<e^{\im( - p_{n-1} + p_n - \ell)} r_n e_1^n 
+ e^{\im( k_n -p_{n-1} + q_n -\ell)} s_n e_2^n| \\
&& + |e_2^{n+1}\>\<- e^{\im(- k_n -q_{n+1} + p_n - \ell)} s_n e_1^n 
+ e^{\im (-q_{n+1} +q_n - \ell)} r_n e_2^n| \\
&=&
\sum_{n \in {\mathbb Z}} |e_1^{n-1} \>\< r_n e_1^n 
+ e^{\im( k_n-n k_1)} s_n e_2^n|
 + |e_2^{n+1}\>\<- e^{\im(- k_n + nk_1)} s_n e_1^n 
+  r_n e_2^n| \\
&=& 
\sum_{n \in {\mathbb Z}} |e_1^{n-1} \>\< r_n e_1^n 
+ e^{\im \theta_n} s_n e_2^n|
 + |e_2^{n+1}\>\<- e^{- \im \theta_n} s_n e_1^n 
+  r_n e_2^n|,
\end{eqnarray*}
where $\theta_n = k_n -n k_1$.
Here, $\theta_0 = \theta_1 =0$.

Consequently, we have the next theorem.


\begin{theorem}\label{thm:1.1}
A one-dimensional quantum walk $U$ is 
unitary equivalent to 
\[
U_{r,\theta} = \sum_{n\in{\mathbb Z}} 
|e_1^{n-1} \>\< r_n e_1^n +  e^{\im\theta_n} s_n e_2^n| 
+ |e_2^{n+1}\>\<  - e^{- \im\theta_n}s_n e_1^n + r_n e_2^n|
\]
for some $0\le r_n \le 1$ and $\theta_n \in {\mathbb R}$ 
with $s_n = \sqrt{1-r_n^2}$ and $\theta_0 = \theta_1 = 0$.
\end{theorem}

The operator $U_{r,\theta}$ is similar to the CMV matrix introduced in \cite{CGMV3, CGMV2}.
However, our approach and result are different in three ways from 
those used in \cite{CGMV3, CGMV2}.
First, our starting point is the general one-dimensional quantum walk.
Second, we add the condition $\theta_0 = \theta_1 =0$.
Third, with the exception of certain special cases,
$U_{r,\theta}$ and $U_{r',\theta'}$ are not unitary equivalent 
if $r\neq r'$ or $\theta \neq \theta'$, as can be seen from the next theorem.



\begin{theorem}\label{thm:1.1.2}
When $0<r_n, r'_n <1$ and $\theta_n, \theta'_n \in [0, 2\pi)$, 
$U_{r,\theta}$ and $U_{r',\theta'}$ are unitary equivalent if and only if
$r= r'$ and $\theta =\theta'$.
\end{theorem}

\proof
We assume that $U_{r,\theta}$ and $U_{r', \theta'}$ are unitary equivalent,
where $0< r_n, r'_n<1$ and $\theta_n, \theta'_n \in [0,2\pi)$ with
$\theta_0=\theta_1= \theta'_0=\theta'_1 =0$.
Then, there exist $\lambda \in {\mathbb R}$ 
and a unitary operator $W =\bigoplus_{n \in {\mathbb Z}} W_n$ 
on $\Hc =\bigoplus_{n \in{\mathbb Z}}\Hc_n$ such that
\[
e^{\im \lambda} WU_{r, \theta} W^* = U_{r', \theta'}.
\]
Then,
\[
e^{\im \lambda} WU_{r,\theta}W^* 
= e^{\im \lambda} \sum_{n \in {\mathbb Z}} |W e_1^{n-1} \>\< r_n W e_1^n 
+ e^{\im \theta_n} s_n W e_2^n|
 + |W e_2^{n+1}\>\<- e^{- \im \theta_n} s_n W e_1^n 
+  r_n W e_2^n| 
\]
and
\[
 U_{r', \theta'} =
\sum_{n \in {\mathbb Z}} |e_1^{n-1} \>\< r'_n e_1^n 
+ e^{\im \theta'_n} s'_n e_2^n|
 + |e_2^{n+1}\>\<- e^{- \im \theta'_n} s'_n e_1^n 
+  r'_n e_2^n|.
\]
Since $P_{n\pm 1} e^{\im \lambda} WU_{r,\theta} W^* P_n = P_{n\pm 1} U_{r',\theta'}P_n$
for all $n \in{\mathbb Z}$,
$We_1^n$ and $We_2^n$ are described as
$We_1^n = e^{\im u_n} e_1^n$ and $We_2^n = e^{\im v_n}e_2^n$ for some 
$u_n, v_n \in {\mathbb R}$.
Then,
\begin{eqnarray}
&& e^{\im \lambda} WU_{r,\theta}W^*  \nonumber \\
&=& \sum_{n \in {\mathbb Z}} | e_1^{n-1} \>\< e^{\im(  - u_{n-1} + u_n-\lambda)} r_n  e_1^n 
+ e^{\im (\theta_n  -u_{n-1} + v_n-\lambda)} s_n  e_2^n|\nonumber \\
&& + |e_2^{n+1}\>\<- e^{\im (-\theta_n  -v_{n+1} + u_n-\lambda)} s_n  e_1^n 
+e^{\im(-v_{n+1} +v_n-\lambda)}  r_n  e_2^n|. \label{eq:WUW}
\end{eqnarray}
Comparing the coefficients of $|e_1^{n-1}\>\<e_1^n|$ and $|e_2^{n+1}\>\<e_2^n|$ yields
\[
 -u_{n-1} +u_n -\lambda = 0, \qquad  - v_{n+1}+v_n - \lambda = 0. 
\]
Here, we can assume that $u_0 = 0$
because $WUW^* = (e^{\im w}W) U (e^{\im w}W)^*$ for any $w \in {\mathbb R}$. 
Therefore, $u_n = n \lambda$.
Moreover, the coefficients of $|e_1^{-1}\>\< e_2^0|$ imply
\[
0 = \theta'_0 = \theta_0 -u_{-1} + v_0 -\lambda
\]
with the result that $v_0 =0$. Hence, $v_n = - n\lambda$.
Similarly, the coefficients of $|e_1^{0}\>\< e_2^1|$ imply
\[
0 = \theta'_1 = \theta_1  -u_0 +v_1-\lambda = -2 \lambda \quad ({\rm mod} \ 2\pi)
\]
with the result that $\lambda=0, \pi$.
When $\lambda =0$, $u_n = v_n =0$, 
and therefore, $W = I_\Hc$ and $U_{r,\theta} = U_{r', \theta'}$.
When $\lambda =\pi$, $u_n = n\pi$ and $v_n =-n\pi = n\pi$ $({\rm mod} \ 2\pi)$.
In this case, $W = \bigoplus_{n \in {\mathbb Z}} (-1)^n I_{\Hc_n}$.
By \eqref{eq:WUW}, we have $e^{\im \lambda}WU_{r,\theta}W^* = U_{r,\theta}$, 
and therefore, $U_{r,\theta} = U_{r', \theta'}$. 
It is easy to see that $U_{r,\theta} = U_{r', \theta'}$ implies that $r=r'$ and $\theta= \theta'$.

The converse is obvious.
\endproof

We next consider some special cases of the one-dimensional quantum walk,
and introduce four types of one-dimensional quantum walk.


\begin{definition}
Let $U$ be a one-dimensional quantum walk expressed by
\[
U =\sum_{n \in {\mathbb Z}} 
|\xi_{n-1,n}\>\< \zeta_{n-1,n}| + |\xi_{n+1,n}\>\<\zeta_{n+1,n}|.
\]

(i)
$U$ is called a translation-invariant quantum walk if there exist 
$\xi_1, \xi_2, \zeta_1, \zeta_2 \in {\mathbb C}^2$ such that
\begin{equation}\label{eq:trans-inv}
\xi_{n,n+1} = \xi_1 , \quad \xi_{n, n-1} = \xi_2 , \quad \zeta_{n-1,n} = \zeta_1, \quad
\zeta_{n+1,n} = \zeta_2 \nonumber
\end{equation}
for all $n \in {\mathbb Z}$.
In other words, the vectors $\xi_{n,n+1}$ are the same, and 
$\xi_{n,n-1}, \zeta_{n-1,n}, \zeta_{n+1,n}$ also satisfy similar conditions.

\medskip
(ii) $U$ is called a one-dimensional quantum walk with one defect if there exist
$\xi_1, \xi_2, \zeta_1, \zeta_2 \in {\mathbb C}^2$ such that
\begin{equation}\label{eq:onedef}
\xi_{n,n+1} = \xi_1 , \quad \xi_{n, n-1} = \xi_2 , \quad \zeta_{n-1,n} = \zeta_1, \quad
\zeta_{n+1,n} = \zeta_2 \nonumber
\end{equation}
for all $n \in {\mathbb Z} \backslash \{ 0 \}$.
In other words, the vectors $\xi_{n,n+1}$ are the same except $n \neq 0$, and 
$\xi_{n,n-1}, \zeta_{n-1,n}, \zeta_{n+1,n}$ also satisfy similar conditions.

\medskip
(iii) \  $U$ is called a complete two-phase quantum walk if there exist
$\xi_1^+, \xi_1^-, \xi_2^+, \xi_2^-, \zeta_1^+, \zeta_1^-, \zeta_2^+, \zeta_2^- \in {\mathbb C}^2$
such that
\begin{equation}\label{eq:2phase1}
\xi_{n,n+1} = \xi_1^+ , \quad \xi_{n, n-1} = \xi_2^+ , \quad \zeta_{n-1,n} = \zeta_1^+, \quad
\zeta_{n+1,n} = \zeta_2^+ \nonumber
\end{equation}
for all $n \ge 0$ and
\begin{equation}\label{eq:2phase2}
\xi_{n,n+1} = \xi_1^- , \quad \xi_{n, n-1} = \xi_2^- , \quad \zeta_{n-1,n} = \zeta_1^-, \quad
\zeta_{n+1,n} = \zeta_2^- \nonumber
\end{equation}
for all $n \le -1$.

\medskip
(iv) $U$ is called a two-phase quantum walk with one defect if there exist
$\xi_1^+, \xi_1^-, \xi_2^+, \xi_2^-, \zeta_1^+, \zeta_1^-, \zeta_2^+, \zeta_2^- \in {\mathbb C}^2$
such that
\begin{equation}\label{eq:2p1d1}
\xi_{n,n+1} = \xi_1^+ , \quad \xi_{n, n-1} = \xi_2^+ , \quad \zeta_{n-1,n} = \zeta_1^+, \quad
\zeta_{n+1,n} = \zeta_2^+
\end{equation}
for all $n \ge 1$ and
\begin{equation}\label{eq:2p1d2}
\xi_{n,n+1} = \xi_1^- , \quad \xi_{n, n-1} = \xi_2^- , \quad \zeta_{n-1,n} = \zeta_1^-, \quad
\zeta_{n+1,n} = \zeta_2^-
\end{equation}
for all $n \le -1$.
\end{definition}

The next theorem describes the unitary equivalent classes of two-phase quantum walks 
with one defect.


\begin{theorem}\label{thm:2p1d}
A two-phase quantum walk $U$ with one defect is unitary equivalent to 
\begin{eqnarray*}
U_{r_+, r_-, r_0, \mu_1, \mu_2, \mu_3}
&=&
|e_1^{-1}\>\<r_0 e_1^0 + e^{\im \mu_1} s_0 e_2^0|
+|e_2^{1}\>\< - e^{\im \mu_2} s_0 e_1^0 + e^{\im (\mu_1 + \mu_2)} r_0 e_2^0| \\
&+&
\sum_{n \ge 1}
|e_1^{n-1}\>\< r_+ e_1^n + s_+ e_2^n|
+ |e_2^{n+1} \>\< -e^{\im \mu_3} s_+ e_1^n + e^{\im \mu_3} r_+ e_2^n| \\
&+&
\sum_{n \le -1}
|e_1^{n-1}\>\< r_- e_1^n + s_- e_2^n|
+ |e_2^{n+1} \>\< -s_- e_1^n + r_- e_2^n| 
\end{eqnarray*}
for some $0 \le r_+, r_-, r_0 \le 1$ and $\mu_1, \mu_2, \mu_3 \in {\mathbb R}$
with $s_\varepsilon = \sqrt{1-r_\varepsilon^2}$ $(\varepsilon = +, -, 0)$.
We write $U_{r_+, r_-, r_0, \mu_1, \mu_2, \mu_3} = U_{r, \mu}$ for short.
\end{theorem}

\proof
Let
\[
U =\sum_{n \in {\mathbb Z}} 
|\xi_{n-1,n}\>\< \zeta_{n-1,n}| + |\xi_{n+1,n}\>\<\zeta_{n+1,n}|
\]
be a two-phase quantum walk with one defect. 
Then, by \eqref{eq:2p1d1} and \eqref{eq:2p1d2}, 
Equation \eqref{eq:1.1.1} can be written as
\begin{eqnarray}
W_1 U W_1^*
&=&
|e_1^{-1} \>\<e^{\im a_0}  r_0 e_1^0 +    e^{\im b_0} s_0 e_2^0| 
+ |e_2^{1}\>\<  e^{\im c_0} s_0 e_1^0 +  e^{\im d_0} r_0  e_2^0| \nonumber \\
&+&
\sum_{n \ge 1}
|e_1^{n-1}\>\< e^{\im a_+} r_+ e_1^n + e^{\im b_+} s_+ e_2^n|
+ |e_2^{n+1} \>\< e^{\im c_+} s_+ e_1^n + e^{\im d_+} r_+ e_2^n| \nonumber \\
&+&
\sum_{n \le -1}
|e_1^{n-1}\>\< e^{\im a_-} r_- e_1^n + e^{\im b_-} s_- e_2^n|
+ |e_2^{n+1} \>\< e^{\im c_-} s_- e_1^n + e^{\im d_-} r_- e_2^n| \label{eq2.7}
\end{eqnarray}
for some $0 \le r_\varepsilon \le 1$ and $a_\varepsilon, b_\varepsilon,
 c_\varepsilon, d_\varepsilon \in {\mathbb R}$ 
with $s_\varepsilon = \sqrt{1-r_\varepsilon^2}$ $(\varepsilon = +,-,0)$.

We then modify Step 2 as follows:

\medskip
\noindent
{\bf Step 2'.}
Let $\ell = (b_-+c_- +\pi)/2$. Define $g_n, h_n \in {\mathbb R}$ by
\[
g_n = 
\left\{ 
\begin{array}{ll}
n(\ell -a_+) & (n\ge 0) \\
n(\ell -a_-) -a_- + a_0 & ( n \le -1)
\end{array}
\right.
\]
and
\[
h_n = 
\left\{ 
\begin{array}{ll}
(n-1)(\ell -a_+) -b_+ + \ell  & (n\ge 1) \\
(n-1)(\ell -a_-) +c_-  +a_0 -a_- -\ell  +\pi  & ( n \le 0)
\end{array}
\right. ,
\]
and a unitary $W_2$ on $\Hc$ by
\[
W_2 = \bigoplus_{n \in{\mathbb Z}} e^{\im g_n} |e_1^n\>\<e_1^n| + 
e^{\im h_n} |e_2^n\>\<e_2^n|.
\]
Then, using $a_\varepsilon + d_\varepsilon+ \pi = b_\varepsilon + c_\varepsilon$,
\begin{eqnarray*}
&& e^{\im \ell}W_2W_1UW_1^*W_2^* \\
&=&
|e_1^{-1} \>\<e^{\im (a_0 -g_{-1} +g_0 - \ell)} r_0 e_1^0 +  e^{\im (b_0-g_{-1} +h_0 - \ell)}  s_0 e_2^0| \\
&& + |e_2^{1}\>\<   e^{\im (c_0 - h_1+g_0-\ell)} s_0e_1^0 +  e^{\im (d_0-h_1+h_0-\ell)} r_0e_2^0| \\
&+&  
\sum_{n \ge 1} 
|e_1^{n-1} \>\<  e^{\im (a_+-g_{n-1}+g_n-\ell)}r_+ e_1^n +  e^{\im (b_+-g_{n-1} + h_n-\ell)} s_+e_2^n| \\
&&
+ |e_2^{n+1}\>\<  e^{\im (c_+-h_{n+1}+g_n -\ell)} s_+
 e_1^n + e^{\im (d_+-h_{n+1}+h_n -\ell)}r_+ e_2^n| \\
&+&  
\sum_{n \le -1} 
|e_1^{n-1} \>\<  e^{\im (a_--g_{n-1}+g_n-\ell)}r_- e_1^n +  e^{\im (b_--g_{n-1} + h_n-\ell)} s_-e_2^n| \\
&&
+ |e_2^{n+1}\>\<  e^{\im (c_--h_{n+1}+g_n -\ell)} s_- e_1^n + e^{\im (d_--h_{n+1}+h_n -\ell)}r_- e_2^n| \\
&=&
|e_1^{-1} \>\<r_0 e_1^0 +   e^{\im (b_0-b_-)}s_0 e_2^0| 
+ |e_2^{1}\>\<  - e^{\im (b_+ -b_- + c_0 - c_-)} s_0 e_1^0 
+  e^{\im (b_0+b_+-2b_- +c_0-c_-)}r_0 e_2^0| \\
&+&  
\sum_{n \ge 1 } 
|e_1^{n-1} \>\< r_+ e_1^n +  s_+ e_2^n|
+ | e_2^{n+1}\>\< - e^{ \im ( b_+ - b_- + c_+-c_-)} s_+ e_1^n +
e^{ \im ( b_+ - b_- + c_+-c_-)} r_+ e_2^n| \\
&+&  
\sum_{n \le -1 }
|e_1^{n-1} \>\< r_- e_1^n + s_- e_2^n|
+ | e_2^{n+1}\>\< - s_- e_1^n + r_- e_2^n| \\
&=&
|e_1^{-1}\>\<r_0 e_1^0 + e^{\im \mu_1 } s_0 e_2^0|
+|e_2^{1}\>\< - e^{\im \mu_2} s_0 e_1^0 + e^{\im (\mu_1 + \mu_2)} r_0 e_2^0| \\
&+&
\sum_{n \ge 1}
|e_1^{n-1}\>\< r_+ e_1^n + s_+ e_2^n|
+ |e_2^{n+1} \>\< - e^{\im \mu_3}s_+ e_1^n +  e^{\im \mu_3 }r_+ e_2^n| \\
&+&
\sum_{n \le -1}
|e_1^{n-1}\>\< r_- e_1^n +s_- e_2^n|
+ |e_2^{n+1} \>\< -s_- e_1^n + r_- e_2^n| \\
&=& U_{r, \mu},
\end{eqnarray*}
where $\mu_1 = b_0 - b_-$, $\mu_2 = b_+ - b_- + c_0 - c_+$ and
$\mu_3 =b_+ - b_- +  c_+ - c_-$.
\endproof

The next theorem considers unitary equivalence between $U_{r,\mu}$ and 
$U_{r', \mu'}$.


\begin{theorem}\label{thm:2p1d2}
When $0 < r_\varepsilon, r'_\varepsilon <1$ $(\varepsilon = +, -, 0)$ 
and $\mu_i, \mu'_i \in [0, 2\pi)$ $(i=1,2,3)$,
$U_{r,\mu}$ and $U_{r', \mu'}$ are unitary equivalent if and only if
$r= r'$, $\mu= \mu'$.
\end{theorem}

\proof
We assume that $U_{r,\mu}$ and 
$U_{r', \mu'}$ are unitary equivalent,
where $0< r_\varepsilon , r'_\varepsilon<1$ and $\mu_i, \mu'_i \in [0,2\pi)$.
Then, there exist $\lambda \in {\mathbb R}$ 
and a unitary operator $W =\bigoplus_{n \in {\mathbb Z}} W_n$ 
on $\Hc =\bigoplus_{n \in{\mathbb Z}}\Hc_n$ such that
\[
e^{\im \lambda} WU_{r,\mu}W^* = U_{r', \mu'}.
\]
Here, 
\begin{eqnarray*}
&& e^{\im \lambda} WU_{r,\mu}W^* \\
&=&
e^{\im \lambda} |W e_1^{-1}\>\<r_0 We_1^0 + e^{\im \mu_1 } s_0 We_2^0|
+e^{\im \lambda}
|We_2^{1}\>\< - e^{\im \mu_2} s_0W e_1^0 + e^{\im (\mu_1 + \mu_2 )} r_0 We_2^0| \\
&+&
e^{\im \lambda}
\sum_{n \ge 1}
|We_1^{n-1}\>\< r_+W e_1^n + s_+ We_2^n|
+ |We_2^{n+1} \>\<  -e^{\im \mu_3 } s_+ We_1^n +  e^{\im \mu_3 } r_+ We_2^n| \\
&+&
e^{\im \lambda} \sum_{n \le -1}
|We_1^{n-1}\>\< r_-W e_1^n +s_- We_2^n|
+ |We_2^{n+1} \>\< -s_- We_1^n +  r_- W e_2^n|  
\end{eqnarray*}
and
\begin{eqnarray*}
U_{r', \mu'}
&=&
|e_1^{-1}\>\<r'_0 e_1^0 + e^{\im \mu'_1 } s'_0 e_2^0|
+|e_2^{1}\>\< - e^{\im \mu'_2} s'_0 e_1^0 + e^{\im (\mu'_1 + \mu'_2 )} r'_0 e_2^0| \\
&+&
\sum_{n \ge 1}
|e_1^{n-1}\>\< r'_+ e_1^n + s'_+ e_2^n|
+ |e_2^{n+1} \>\<- e^{\im \mu'_3 }s'_+ e_1^n +e^{\im \mu'_3 } r'_+ e_2^n| \\
&+&
\sum_{n \le -1}
|e_1^{n-1}\>\< r'_- e_1^n + s'_- e_2^n|
+ |e_2^{n+1} \>\< -s'_- e_1^n + r'_- e_2^n|.
\end{eqnarray*}
Considering $ P_{n\pm 1} e^{\im \lambda} WU_{r,\mu}W^* P_{n } = P_{n\pm 1} U_{r', \mu'} P_{n}$ for
any $n \in {\mathbb Z}$,
we have
$We_1^n = e^{\im u_n} e_1^n$ and $We_2^n = e^{\im v_n}e_2^n$ for some 
$u_n, v_n \in {\mathbb R}$.
Then,
\begin{eqnarray}
&& e^{\im \lambda} WU_{r,\mu}W^*  \nonumber \\
&=&
|e_1^{-1} \>\<e^{\im (-u_{-1} + u_0 -\lambda )} r_0 e_1^0 
+  e^{\im (-u_{-1} + v_0 - \lambda +\mu_1)}  s_0 e_2^0|  \nonumber\\
&& + |e_2^{1}\>\< -  e^{\im (-v_1 + u_0 - \lambda +\mu_2)} s_0e_1^0 
+  e^{\im (-v_1 +v_0 -\lambda +\mu_1+\mu_2 )} r_0e_2^0|  \nonumber\\
&+&  
\sum_{n \ge 1} 
|e_1^{n-1} \>\<  e^{\im (-u_{n-1} +u_n -\lambda )}r_+ e_1^n 
+  e^{\im (-u_{n-1} + v_{n}-\lambda)} s_+e_2^n|  \nonumber\\
&&
+ |e_2^{n+1}\>\< - e^{\im (-v_{n+1} +u_n -\lambda +\mu_3)} s_+ e_1^n 
+ e^{\im (-v_{n+1} + v_n -\lambda +\mu_3 )}r_+ e_2^n|  \nonumber\\
&+&  
\sum_{n \le -1} 
|e_1^{n-1} \>\<  e^{\im (-u_{n-1} + u_n-\lambda)}r_- e_1^n 
+  e^{\im (-u_{n-1} + v_n-\lambda)} s_-e_2^n| \nonumber \\ 
&&
+ |e_2^{n+1}\>\< - e^{\im (-v_{n+1} + u_n -\lambda)} s_- e_1^n 
+ e^{\im (-v_{n+1} + v_n-\lambda )}r_- e_2^n|. \label{eq:1.5.1}
\end{eqnarray}
Since $e^{\im \lambda} WU_{r,\mu}W^* = U_{r', \mu'}$, we obtain $r =r'$.
Moreover, comparing the coefficients of $|e_1^{n-1}\>\<e_1^n|$,
$|e_1^{n-1}\>\<e_2^{n}|$ and $|e_2^{n+1}\>\<e_2^n|$
yields 
\begin{equation}\label{eq:1.6.1}
-u_{n-1} +u_n -\lambda = 0, \quad -u_{n-1} + v_{n} -\lambda = 0 \ (n\neq 0), \quad
-v_{n+1} + v_n - \lambda  =0 \ (n \le -1).
\end{equation}
Here, we can assume that $u_0 = 0$, 
because $WUW^* = (e^{\im w}W) U (e^{\im w}W)^*$ for any $w \in {\mathbb R}$. 
Therefore, $u_n = n \lambda$, and this implies $v_{n} = u_{n-1}+\lambda= n \lambda$ $(n \neq 0)$.
Using the third equation in \eqref{eq:1.6.1}, we have $2\lambda = 0$, 
and therefore $\lambda = 0$ or $\pi$.
Moreover, the coefficients of $|e_2^{0}\>\<e_1^{-1}|$ imply
 $ v_0 = u_{-1} - \lambda = -2\lambda =0$.  
Comparing the coefficients of 
$|e_1^{-1}\>\<e_2^0|$, $|e_2^{1}\>\<e_1^0|$
and $|e_2^{n+1}\>\<e_2^n|$ $(n \ge 1)$, we obtain $\mu = \mu'$.
\endproof


From the above proof,
if $e^{\im \lambda} WU_{r,\mu}W^* = U_{r, \mu}$, then $\lambda = 0$ or $\pi$.
When $\lambda = 0$, $v_n = u_n = 0$ and $W = I_{\Hc}$. 
When $\lambda = \pi$, $v_n = u_n = n \pi$ and 
$W = \bigoplus_{n\in{\mathbb Z}} (-1)^n I_{\Hc_n}$.
Hence, we have the next corollary.

\begin{corollary}\label{cor:1.6}
Let $0< r_\varepsilon < 1$ $(\varepsilon = +, - , 0)$ 
and $\mu_i \in [0, 2\pi)$, and let $W = \bigoplus_{n\in{\mathbb Z}}W_n$
be a unitary on $\Hc = \bigoplus_{n\in{\mathbb Z}}\Hc_n$.
Then, for $\lambda \in [0, 2\pi)$, 
\[
e^{\im \lambda} WU_{r,\mu}W^* = U_{r, \mu}
\]
if and only if $\lambda =0$ and $W = I_\Hc$ or 
$\lambda = \pi$ and $W = \bigoplus_{n \in {\mathbb Z}} (-1)^n I_{\Hc_n}$.
\end{corollary}

As a corollary of Theorem \ref{thm:2p1d} and \ref{thm:2p1d2},
we obtain the following.


\begin{corollary}{\rm \cite{GKD}}\label{thm:1.2}
A translation-invariant quantum walk $U$ 
is unitary equivalent to 
\[
U_r= \sum_{n\in{\mathbb Z}}
|e_1^{n-1} \>\< r e_1^n +  s e_2^n| 
+ |e_2^{n+1}\>\<  s e_1^n + r e_2^n|
\]
for some $0 \le r \le 1$ with $s = \sqrt{1-r^2}$.
Moreover, $U_r$ and $U_{r'}$ are unitary equivalent if and only if $r=r'$.
\end{corollary}

\proof
From the definition of a translation-invariant quantum walk,
we can assume that, in \eqref{eq2.7}, $r_+ = r_- =r_0$, $a_+ = a_- =a_0$ and so on.
This implies that $\mu_1 = b_0 - b_- =0$, $\mu_2= b_+-b_- + c_0 - c_+ =0$ and 
$ \mu_3 =b_+ - b_- + c_+ -c_- =0$.
Setting $r = r_0$ satisfies the first assertion.
The necessary and sufficient condition for unitary equivalence 
follows from Theorem \ref{thm:2p1d2}.
\endproof

\begin{corollary}\label{cor:2.11}
A one-dimensional quantum walk $U$ with one defect
is unitary equivalent to 
\begin{eqnarray*}
U_{r_\pm, r_0, \nu_1, \nu_2} &=& 
|e_1^{-1}\>\<r_0 e_1^0 + e^{\im \nu_1} s_0 e_2^0|
+|e_2^{1}\>\< - e^{\im \nu_2} s_0 e_1^0 + e^{\im (\nu_1 + \nu_2  )} r_0 e_2^0| \\
&+& \sum_{n\in{\mathbb Z}\backslash \{0\} }
|e_1^{n-1} \>\< r_\pm e_1^n +  s_\pm e_2^n| 
+ |e_2^{n+1}\>\<  s_\pm e_1^n + r_\pm e_2^n| 
\end{eqnarray*}
for some $0 \le r_\varepsilon \le 1$ and $\nu_1, \nu_2 \in {\mathbb R}$
with $s_\varepsilon = \sqrt{1-r_\varepsilon^2}$
$(\varepsilon = \pm, 0)$.
We write $U_{r_\pm, r_0, \nu_1, \nu_2} = U_{r, \nu}$ for short.
Moreover, when $0 < r_\varepsilon, r'_\varepsilon <1$ and $\nu_i, \nu'_i \in [0,2\pi)$,
$U_{r, \nu} $ and $U_{r',\nu'}$ are unitary equivalent if and only if $r=r'$ and $\nu = \nu'$.
\end{corollary}

\proof
From the definition of a one-dimensional quantum walk with one defect,
we can assume that, in \eqref{eq2.7}, $r_+ =r_-$, $a_+  =a_-$ and so on.
This implies that $\mu_3 =b_+ - b_- + c_+ -c_- =0$.
Setting $r_\pm = r_+ , \nu_1= \mu_1$ and $\nu_2= \mu_2$ satisfies the first assertion.
The necessary and sufficient condition for unitary equivalence 
follows from Theorem \ref{thm:2p1d2}.
\endproof

Clearly, Theorem \ref{thm:2p1d} can be applied to complete two-phase quantum walks,
though in this case, 
 $U_{r, \mu}$ is not a complete two-phase quantum walk.
Indeed, from the definition of complete two-phase quantum walks,
we can assume that, in \eqref{eq2.7}, $r_0 = r_+$, $a_0 = a_+$ and so on.
Then, $\mu_1 = b_0 - b_- \neq 0$ in general,
and the coefficients of $|e_1^{-1}\>\< e_2^0|$ and $|e_1^{n-1}\>\<e_2^n|$ $(n \ge 1)$ 
of $U_{r, \mu}$ are not the same.

Hence, we provide the next theorem.


\begin{theorem}
A complete two-phase quantum walk $U$ is unitary equivalent to 
\begin{eqnarray*}
U_{r_+, r_-, \sigma_1, \sigma_2}
&=&
\sum_{n \ge 0}
|e_1^{n-1}\>\< r_+ e_1^n + s_+ e_2^n|
+ |e_2^{n+1} \>\< -e^{\im \sigma_1} s_+ e_1^n +e^{\im \sigma_1}  r_+ e_2^n| \\
&+&
\sum_{n \le -1}
|e_1^{n-1}\>\< r_- e_1^n + e^{\im \sigma_2}s_- e_2^n|
+ |e_2^{n+1} \>\< -s_- e_1^n + e^{\im \sigma_2} r_- e_2^n| 
\end{eqnarray*}
for some $0 \le r_+, r_- \le 1$ 
and $\sigma_1, \sigma_2 \in {\mathbb R}$
with $s_\varepsilon = \sqrt{1-r_\varepsilon^2}$ $(\varepsilon = +, -)$.
We write $U_{r_+, r_-, \sigma_1, \sigma_2} = U_{r, \sigma}$ for short.
Moreover, when $0 < r_\varepsilon, r'_\varepsilon <1$ and $\sigma_i, \sigma'_i \in [0,2\pi)$,
$U_{r, \sigma} $ and $U_{r',\sigma'}$ are unitary equivalent if and only if 
$r=r'$ and $\sigma = \sigma'$.
\end{theorem}

\proof
The proof is almost the same as that given for Theorems \ref{thm:2p1d} and \ref{thm:2p1d2},
but we present it here for completeness.
Let
\[
U =\sum_{n \in {\mathbb Z}} 
|\xi_{n-1,n}\>\< \zeta_{n-1,n}| + |\xi_{n+1,n}\>\<\zeta_{n+1,n}|
\]
be a complete two-phase quantum walk.
Then, by definition,
Equation \eqref{eq:1.1.1} can be written as
\begin{eqnarray}
W_1 U W_1^*
&=&
\sum_{n \ge 0}
|e_1^{n-1}\>\< e^{\im a_+} r_+ e_1^n + e^{\im b_+} s_+ e_2^n|
+ |e_2^{n+1} \>\< e^{\im c_+} s_+ e_1^n + e^{\im d_+} r_+ e_2^n| \nonumber \\
&+&
\sum_{n \le -1}
|e_1^{n-1}\>\< e^{\im a_-} r_- e_1^n + e^{\im b_-} s_- e_2^n|
+ |e_2^{n+1} \>\< e^{\im c_-} s_- e_1^n + e^{\im d_-} r_- e_2^n| \nonumber
\end{eqnarray}
for some $0 \le r_\varepsilon \le 1$ and $a_\varepsilon, b_\varepsilon,
 c_\varepsilon, d_\varepsilon \in {\mathbb R}$ 
with $s_\varepsilon = \sqrt{1-r_\varepsilon^2}$ $(\varepsilon = +,-)$.

We then modify Step 2 as follows:

\medskip
\noindent
{\bf Step 2''.}
Let $\ell = (b_++c_- +\pi)/2$. Define $g_n, h_n \in {\mathbb R}$ by
\[
g_n = 
\left\{ 
\begin{array}{ll}
n(\ell -a_+) & (n\ge 0) \\
n(\ell -a_-) -a_- + a_+ & ( n \le -1)
\end{array}
\right.
\]
and
\[
h_n = 
\left\{ 
\begin{array}{ll}
(n-1)(\ell -a_+) -b_+ + \ell  & (n\ge 1) \\
(n-1)(\ell -a_-) +c_-  +a_+ -a_- -\ell  +\pi  & ( n \le 0)
\end{array}
\right. ,
\]
and a unitary $W_2$ on $\Hc$ by
\[
W_2 = \bigoplus_{n \in{\mathbb Z}} e^{\im g_n} |e_1^n\>\<e_1^n| + 
e^{\im h_n} |e_2^n\>\<e_2^n|.
\]
Then, using $a_\varepsilon + d_\varepsilon+ \pi = b_\varepsilon + c_\varepsilon$,
\begin{eqnarray*}
&& e^{\im \ell}W_2W_1UW_1^*W_2^* \\
&=&
\sum_{n \ge 0} 
|e_1^{n-1} \>\<  e^{\im (a_+-g_{n-1}+g_n-\ell)}r_+ e_1^n +  e^{\im (b_+-g_{n-1} + h_n-\ell)} s_+e_2^n| \\
&&
+ |e_2^{n+1}\>\<  e^{\im (c_+-h_{n+1}+g_n -\ell)} s_+
 e_1^n + e^{\im (d_+-h_{n+1}+h_n -\ell)}r_+ e_2^n| \\
&+&  
\sum_{n \le -1} 
|e_1^{n-1} \>\<  e^{\im (a_--g_{n-1}+g_n-\ell)}r_- e_1^n +  e^{\im (b_--g_{n-1} + h_n-\ell)} s_-e_2^n| \\
&&
+ |e_2^{n+1}\>\<  e^{\im (c_--h_{n+1}+g_n -\ell)} s_- e_1^n + e^{\im (d_--h_{n+1}+h_n -\ell)}r_- e_2^n| \\
&=&
\sum_{n \ge 0 } 
|e_1^{n-1} \>\< r_+ e_1^n +  s_+ e_2^n|
+ | e_2^{n+1}\>\< - e^{ \im (  c_+-c_-)} s_+ e_1^n +
e^{ \im (  c_+-c_-)} r_+ e_2^n| \\
&+&  
\sum_{n \le -1 }
|e_1^{n-1} \>\< r_- e_1^n + e^{\im(b_- - b_+)} s_- e_2^n|
+ | e_2^{n+1}\>\< - s_- e_1^n +e^{\im (b_- - b_+)} r_- e_2^n| \\
&=&
\sum_{n \ge 0}
|e_1^{n-1}\>\< r_+ e_1^n +   s_+ e_2^n|
+ |e_2^{n+1} \>\< - e^{\im \sigma_1}s_+ e_1^n +  e^{\im \sigma_1 }r_+ e_2^n| \\
&+&
\sum_{n \le -1}
|e_1^{n-1}\>\< r_- e_1^n + e^{\im \sigma_2}s_- e_2^n|
+ |e_2^{n+1} \>\< -s_- e_1^n +  e^{\im \sigma_2}r_- e_2^n| \\
&=& U_{r, \sigma},
\end{eqnarray*}
where $\sigma_1 = c_+- c_-$ and $\sigma_2 = b_- - b_+$.
This shows the first assertion of this theorem.

Next, assume that $U_{r,\sigma}$ and 
$U_{r', \sigma'}$ are unitary equivalent,
where $0< r_\varepsilon , r'_\varepsilon<1$ and $\sigma_i, \sigma'_i \in [0,2\pi)$.
Then, there exist $\lambda \in {\mathbb R}$ 
and a unitary operator $W =\bigoplus_{n \in {\mathbb Z}} W_n$ 
on $\Hc =\bigoplus_{n \in{\mathbb Z}}\Hc_n$ such that
\[
e^{\im \lambda} WU_{r,\sigma}W^* = U_{r',\sigma'}.
\]
Here, 
\begin{eqnarray*}
&& e^{\im \lambda} WU_{r,\sigma}W^* \\
&=&
e^{\im \lambda}
\sum_{n \ge 0}
|We_1^{n-1}\>\< r_+ We_1^n +   s_+ We_2^n|
+ |We_2^{n+1} \>\< - e^{\im \sigma_1}s_+ We_1^n +  e^{\im \sigma_1 }r_+ We_2^n| \\
&+&
\sum_{n \le -1}
|We_1^{n-1}\>\< r_- W e_1^n + e^{\im \sigma_2}s_- We_2^n|
+ |We_2^{n+1} \>\< -s_-W e_1^n +  e^{\im \sigma_2}r_- We_2^n| 
\end{eqnarray*}
and
\begin{eqnarray*}
U_{r', \sigma'}
&=&
\sum_{n \ge 0}
|e_1^{n-1}\>\< r'_+ e_1^n +   s'_+ e_2^n|
+ |e_2^{n+1} \>\< - e^{\im \sigma'_1}s'_+ e_1^n +  e^{\im \sigma'_1 }r'_+ e_2^n| \\
&+&
\sum_{n \le -1}
|e_1^{n-1}\>\< r'_- e_1^n + e^{\im \sigma'_2}s'_- e_2^n|
+ |e_2^{n+1} \>\< -s'_- e_1^n +  e^{\im \sigma'_2}r'_- e_2^n| .
\end{eqnarray*}
Considering $ P_{n\pm 1} e^{\im \lambda} WU_{r,\sigma}W^* P_{n } = P_{n\pm 1} U_{r', \sigma'} P_{n}$ for
any $n \in {\mathbb Z}$,
we have
$We_1^n = e^{\im u_n} e_1^n$ and $We_2^n = e^{\im v_n}e_2^n$ for some 
$u_n, v_n \in {\mathbb R}$.
Then,
\begin{eqnarray}
&& e^{\im \lambda} WU_{r,\sigma}W^*  \nonumber \\
&=&
\sum_{n \ge 0} 
|e_1^{n-1} \>\<  e^{\im (-u_{n-1} +u_n -\lambda )}r_+ e_1^n 
+  e^{\im (-u_{n-1} + v_{n}-\lambda)} s_+e_2^n|  \nonumber\\
&&
+ |e_2^{n+1}\>\< - e^{\im (-v_{n+1} +u_n -\lambda +\sigma_1)} s_+ e_1^n 
+ e^{\im (-v_{n+1} + v_n -\lambda +\sigma_1 )}r_+ e_2^n|  \nonumber\\
&+&  
\sum_{n \le -1} 
|e_1^{n-1} \>\<  e^{\im (-u_{n-1} + u_n-\lambda)}r_- e_1^n 
+  e^{\im (-u_{n-1} + v_n-\lambda+\sigma_2)} s_-e_2^n| \nonumber \\ 
&&
+ |e_2^{n+1}\>\< - e^{\im (-v_{n+1} + u_n -\lambda)} s_- e_1^n 
+ e^{\im (-v_{n+1} + v_n-\lambda +\sigma_2)}r_- e_2^n|. \nonumber
\end{eqnarray}
Since $e^{\im \lambda} WU_{r,\sigma}W^* = U_{r', \sigma'}$, we obtain $r =r'$.
Moreover, comparing the coefficients of $|e_1^{n-1}\>\<e_1^n|$,
$|e_1^{n-1}\>\<e_2^{n}|$ and $|e_2^{n+1}\>\<e_1^n|$
yields
\begin{equation}
-u_{n-1} +u_n -\lambda = 0, \quad -u_{n-1} + v_{n} -\lambda = 0 \ (n\ge 0), \quad
-v_{n+1} + u_n - \lambda  =0 \ (n \le -1). \nonumber
\end{equation}
Here, we can assume that $u_0 = 0$, 
because $WUW^* = (e^{\im w}W) U (e^{\im w}W)^*$ for any $w \in {\mathbb R}$. 
Therefore, $u_n = n \lambda$, and this implies 
$v_n = u_{n-1} +\lambda = n \lambda$ $(n \ge 0)$ and 
$v_{n+1} = u_n - \lambda = (n-1) \lambda$ $(n \le -1)$.
Hence, $v_0 = 0$ and $v_0 = -2\lambda$ with the result that $\lambda = 0$ or $\pi$.
Then, $-v_{n+1} + v_n - \lambda = 0$ for all $n \in {\mathbb Z}$, and therefore,
$\sigma = \sigma'$, comparing the coefficients of $|e_2^{n+1}\>\<e_2^n|$.
This completes the proof.
\endproof

From the above proof,
if $e^{\im \lambda} WU_{r,\sigma}W^* = U_{r, \sigma}$, then $\lambda = 0$ or $\pi$.
When $\lambda = 0$, $v_n = u_n = 0$ and $W = I_{\Hc}$. 
When $\lambda = \pi$, $v_n = u_n = n \pi$ and 
$W = \bigoplus_{n\in{\mathbb Z}} (-1)^n I_{\Hc_n}$.
Hence, we have the next corollary.

\begin{corollary}
Let $0< r_\varepsilon < 1$ $(\varepsilon = +, -)$ 
and $\sigma_i \in [0, 2\pi)$ $(i=1,2)$, and let $W = \bigoplus_{n\in{\mathbb Z}}W_n$
be a unitary on $\Hc = \bigoplus_{n\in{\mathbb Z}}\Hc_n$.
Then, for $\lambda \in [0, 2\pi)$, 
\[
e^{\im \lambda} WU_{r,\sigma}W^* = U_{r, \sigma}
\]
if and only if $\lambda =0$ and $W = I_\Hc$ or 
$\lambda = \pi$ and $W = \bigoplus_{n \in {\mathbb Z}} (-1)^n I_{\Hc_n}$.
\end{corollary}


\section{Unitary equivalent classes of one-dimensional quantum walks with initial states}\label{sec3}

In this section, we consider one-dimensional quantum walks with initial states.
We assume that an initial state $\Phi$ is in $\Hc_0$.

\begin{definition}
One-dimensional quantum walks with initial states 
$(U,  \Phi)$ and 
$(U' , \Phi')$
are unitary equivalent if there exists a unitary $W = \bigoplus_{n\in{\mathbb Z}} W_n$ on
$\Hc = \bigoplus_{n\in{\mathbb Z}} \Hc_n$ such that
\[
U' = WUW^* \quad {\rm and} \quad \Phi' = W \Phi.
\]
\end{definition}

Unitary equivalent classes of two-phase quantum walks with one defect with initial 
states are described as follows:

\begin{theorem}\label{thm:2.2}
A two-phase quantum walk with one defect with an initial state $(U, \Phi)$ is 
unitary equivalent to $(U_{r, \mu}, \Phi_{\alpha, \theta})$ for some
$0 \le r_\varepsilon, \alpha \le 1$ $(\varepsilon = +, -, 0)$, $\mu_i, \theta \in {\mathbb R}$ $(i =1,2,3)$,
where  $\Phi_{\alpha, \theta} = \alpha e_1^0 +e^{\im \theta} \sqrt{1-\alpha^2} e_2^0$.

Moreover, $(U_{r, \mu} ,  \Phi_{\alpha, \theta})$ and 
$(U_{r',\mu'} , \Phi_{\alpha', \theta'})$ 
with $ 0 < r_\varepsilon, r'_\varepsilon, \alpha, \alpha' <1$ and $\mu_i, \mu'_i, \theta, \theta' \in [0, 2\pi)$ 
are unitary equivalent
if and only if $r=r'$, $\mu= \mu'$, $\alpha=\alpha'$ and $\theta = \theta'$.
\end{theorem}

\proof
It was proved that $U$ is unitary equivalent to $U_{r,\mu}$ for some $r$ and $\mu$
in Theorem \ref{thm:2p1d}. 
Hence, there exists a unitary $W = \bigoplus_{n\in{\mathbb Z}} W_n$ 
on $\Hc$ such that $WUW^* = U_{r,\mu}$.
The state $W \Phi \in \Hc_0 ={\mathbb C}^2$ can be written as
$W \Phi = \alpha e_1^0 + \beta e_2^0$ for some $\alpha ,\beta \in {\mathbb C}$ 
with $|\alpha|^2 + |\beta|^2 =1$.
Since $W\Phi$ and $e^{\im \lambda} W\Phi$ are identified, we can assume that
$0 \le \alpha \le 1$.
Then, $\beta = e^{\im \theta} \sqrt{1-\alpha^2}$ 
for some $\theta \in {\mathbb R}$.
Therefore, $W \Phi = \Phi_{\alpha, \theta}$ holds, and hence, $(U, \Phi)$ is unitary equivalent
to $(U_{r,\mu}, \Phi_{\alpha, \theta})$.

Next, assume that 
 $(U_{r,\mu},  \Phi_{\alpha, \theta})$ and 
$(U_{r',\mu'} , \Phi_{\alpha', \theta'})$ 
with $ 0 < r_\varepsilon, r'_\varepsilon ,\alpha, \alpha'  <1$ 
and $\mu_i, \mu'_i, \theta, \theta' \in [0, 2\pi)$ 
 are unitary equivalent.
Then, by Theorem \ref{thm:2p1d2}, $r= r'$ and $\mu = \mu'$.
Moreover, if there exist $\lambda \in {\mathbb R}$ and 
a unitary operator $W = \bigoplus_{n\in{\mathbb Z}} W_n$ on $\Hc$ such that
\[
e^{\im\lambda} W U_{r, \mu} W^* = U_{r,\mu},
\]
then, by Corollary \ref{cor:1.6}, $\lambda = 0$ and $W=I$ or $\lambda= \pi$ and
$W =\bigoplus_{n\in{\mathbb Z}} (-1)^n I_{\Hc_n}$.
Therefore, $W\Phi_{\alpha, \theta} = \Phi_{\alpha', \theta'}$ implies 
 $\alpha=\alpha'$ and $\theta = \theta'$.
\endproof

As a corollaries, and from Corollaries \ref{thm:1.2} and \ref{cor:2.11},
we have the following.

\begin{corollary}
A translation-invariant quantum walk $(U,  \Phi)$ is unitary equivalent
to $(U_r, \Phi_{\alpha, \theta})$ for some $0\le r , \alpha \le 1$ and 
$\theta \in {\mathbb R}$.

Moreover, $(U_r, \Phi_{\alpha, \theta})$ and 
$(U_{r'} , \Phi_{\alpha', \theta'})$ 
with $0 < r,r',\alpha , \alpha' < 1$ and $\theta, \theta' \in [0, 2\pi)$ are unitary equivalent
if and only if $r=r'$, $\alpha=\alpha'$ and $\theta = \theta'$.
\end{corollary}

\begin{corollary}
A one-dimensional quantum walk with one defect $(U,  \Phi)$ is unitary equivalent
to $(U_{r,\nu}, \Phi_{\alpha, \theta})$ for some $0\le r_\varepsilon , \alpha \le 1$
$(\varepsilon = \pm , 0)$ and 
$\nu_i, \theta \in {\mathbb R}$ $(i=1,2)$.

Moreover, $(U_{r, \nu}, \Phi_{\alpha, \theta})$ and 
$(U_{r', \nu'} , \Phi_{\alpha', \theta'})$ 
with $0< r_\varepsilon, r'_\varepsilon ,\alpha, \alpha' <1$ 
and $\nu_i, \nu'_i, \theta, \theta' \in [0, 2\pi)$ are unitary equivalent
if and only if $r=r'$, $\nu =\nu'$, $\alpha=\alpha'$ and $\theta = \theta'$.
\end{corollary}

The proof of the next theorem is almost the same as that given for Theorem \ref{thm:2.2}
and is omitted.

\begin{theorem}
A complete two-phase quantum walk $(U,  \Phi)$ is unitary equivalent
to $(U_{r,\sigma}, \Phi_{\alpha, \theta})$ for some 
$0\le r_\varepsilon , \alpha \le 1$ $(\varepsilon = +,-)$ and 
$\sigma_i, \theta \in {\mathbb R}$ $(i=1,2)$.

Moreover, $(U_{r, \sigma}, \Phi_{\alpha, \theta})$ and 
$(U_{r',\sigma'} , \Phi_{\alpha', \theta'})$ 
with $0< r_\varepsilon, r'_\varepsilon, \alpha , \alpha' <1$ and 
$\sigma_i, \sigma'_i, \theta, \theta' \in [0, 2\pi)$ are unitary equivalent
if and only if $r=r'$, $\sigma =\sigma'$, $\alpha=\alpha'$ and $\theta = \theta'$.
\end{theorem}


\end{document}